\documentclass[aps,showpacs,preprint]{revtex4}
\usepackage{amsmath}
\usepackage{amssymb}
\usepackage{amscd}
\usepackage{wasysym}
\usepackage[ansinew]{inputenc}
\usepackage[T1]{fontenc}
\usepackage{ae,aecompl}

\usepackage[dvips]{graphicx}
\DeclareGraphicsExtensions{.eps}

\newcommand{\be}{\begin{equation}}
\newcommand{\ee}{\end{equation}}
\newcommand{\bea}{\begin{eqnarray}}
\newcommand{\eea}{\end{eqnarray}}
\newcommand{\nn}{\nonumber}
\newcommand{\ket}[1]{|#1\rangle}

\begin{document}
\bibliographystyle{apsrev}

\title{Quantum walks and quantum simulations with Bloch oscillating spinor atoms}
\author{D. Witthaut}
\affiliation{QUANTOP, The Niels Bohr Institute, University of Copenhagen,
 		DK--2100 Copenhagen, Denmark}  
\affiliation{Network Dynamics Group,
Max-Planck-Institute for Dynamics and Self-Organization,
D--37073 G\"ottingen, Germany}
\date{\today }

\begin{abstract}
We propose a scheme for the realization of a quantum walker and
a quantum simulator for the Dirac equation with ultracold spinor atoms 
in driven optical lattices. A precise control of the dynamics of the atomic
matter wave can be realized using time-dependent external forces.
If the force depends on the spin state of the atoms, the dynamics will 
entangle the inner and outer degrees of freedom which offers unique
opportunities for quantum information and quantum simulation.
Here, we introduce a method to realize a quantum walker based on the 
state-dependent transport of spinor atoms and a coherent driving
of the internal state. 
In the limit of weak driving the dynamics is equivalent to that of a Dirac 
particle in 1+1 dimensions. Thus it becomes possible to simulate 
relativistic effects such as Zitterbewegung and Klein tunneling.
\end{abstract}

\maketitle

\section{Introduction}

Quantum simulators aim at the simulation of complex quantum
systems in well controllable laboratory experiments \cite{Bulu09}.
Such a simulation is especially useful
when the original quantum system is experimentally not accessible
and classical simulations are impossible due to the exponential
size of the Hilbert space. 
Furthermore, quantum simulators offer the possibility
to tune the experimental parameters to explore novel 
physical phenomena. Important examples include the
simulation of solid state systems with ultracold atoms \cite{Bloc08},
Dirac dynamics with graphene \cite{Novo05,Kats06} or trapped ions 
\cite{Lama07,Gerr10} and sonic black holes in Bose-Einstein 
condensates \cite{Gara00}.

Ultracold atoms in optical lattices are especially suited for such 
a task, since their dynamics can be controlled with an astonishing
precision and their dynamics can be measured in situ.
In the present paper we propose to use spinor atoms in
tilted or driven optical lattices to realize a quantum walker --
a paradigmatic system in quantum information science 
\cite{Ahar93,Kars09,Schm09,Schr10,Zahr10} -- 
and a quantum simulator for relativistic 
Dirac dynamics. In contrast to previous proposals which were based 
on the realization of artificial gauge fields using atoms with a tripod
internal structure \cite{Zhu07,Otte09}, we focus on simple spinor atoms with 
two internal states, which are used routinely in ongoing 
experiments. The necessary correlations between 
the internal and the external dynamics can be realized with
state dependent external forces, in particular by
magnetic gradient fields \cite{Gorl03,Gust08,Weld09,Bohi09,Hall10}. 
Using a suitable driving of this
external field and microwave transitions between the internal
states \cite{Fors09,Misc10}, one can obtain control over the full 
dynamics of the atoms.

\section{Bloch oscillations and transport in optical lattices}

\begin{figure}[t]
\centering
\includegraphics[width=10cm,  angle=0]{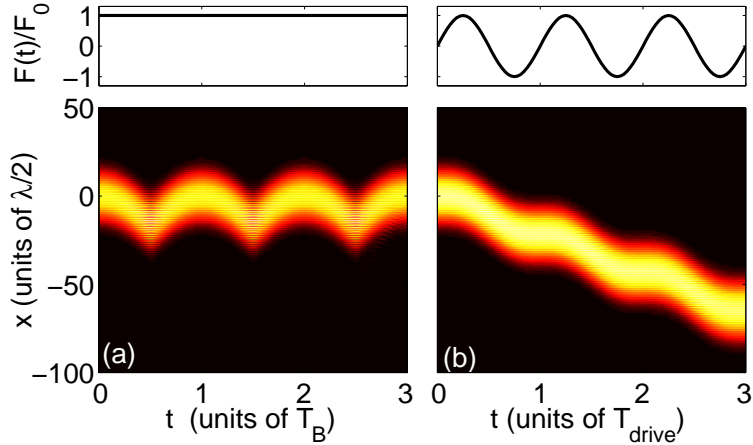}
\caption{\label{fig-flipped1}  
(Color online)
Dynamics of ultracold Cesium atoms in a driven optical lattice:
(a) A static external field leads to Bloch oscillations.
(b) A sinusoidal driving of the field leads to directed transport.
We assumed an optical lattice with a wave length 
$\lambda = 1064 \, {\rm nm}$, 
a depth of $V_0 =  E_R$ and a maximum field strength of
$F_0/m = 0.42 \, {\rm m/s}^2$. The time is given in units of the 
Bloch period $T_B = 8.44 \, {\rm ms}$ and the period of the
external driving $T_{\rm drive} = 10 \, {\rm ms}$, respectively.
}
\end{figure}

A Bloch oscillation is the counter-intuitive dynamics of a quantum
particle in a periodic potential subject to a static external force.
The force accelerates the particle until it reaches the edge of 
the Brillouin zone, where it is Bragg reflected. If Landau-Zener 
tunneling to higher bands can be neglected, this leads to a fully
periodic motion. This peculiar kind of dynamics was postulated 
by Bloch already in 1928 in the context of electrons in crystals 
\cite{Bloc28} but never observed because of the strong scattering 
of the electrons. Thus it took another seven decades before
Bloch oscillations could be demonstrated for electrons in
semiconductor superlattices \cite{Feld92}, photons in
waveguide arrays \cite{Pert99} and ultracold atoms in 
optical lattices \cite{Daha96}. Especially the latter realization 
shows an astonishing level of coherence and offers various 
possibilities to precisely control the atomic dynamics. Recent
experiments demonstrated Bloch oscillations over tenth of 
thousands of periods and macroscopic distances \cite{Gust08}.
An example of Bloch oscillations of an atomic matter wave
is shown in Fig.~\ref{fig-flipped1} (a).

Extensive possibilities to control the atomic motion can
be realized in driven optical lattices, i.e. in lattices with a 
time-dependent external field 
\cite{Hart04,Klum07,Brei07,Brei08,Sias08,Zene09,Hall10}.
A particular interesting case is a periodic driving, when
the direction of the field is reversed before the matter
wave reaches the edge of the Brillouin zone,
decelerating the wave packet back to a standstill. 
Thus the atoms always have a positive momentum 
such that directed transport is realized, as illustrated
in Fig.~\ref{fig-flipped1} (b) for a sinusoidal driving.

To be precise, we have simulated the dynamics of ultracold
Cs atoms in an optical lattice with a wavelength of
$\lambda = 1064 \, {\rm nm}$ and a depth of $V_0 = E_R$.
The strength of the external field was assumed to be
$F_0/m = 0.42 \, {\rm m/s}^2$, which can be easily realized by 
accelerating the optical lattice \cite{Daha96} or by a magnetic 
gradient field \cite{Gust08,Hall10}. 
The dynamics of the atomic matter wave is then given by 
the Wannier-Stark Hamiltonian
\be
  \hat H(t) =  \frac{-\hbar^2}{2m}
    \frac{\partial^2 }{\partial x^2} + V_0 \cos^2(k_0 x)  + F(t) x,
\ee
where $k_0 = 2\pi/\lambda$ and $E_R = \hbar^2 k_0^2/2m$
is the atomic recoil energy. The initial state has been chosen
as a Bloch state with zero momentum weighted by a Gaussian
with a width $\sigma = 6 \, \lambda$.
For a static field one then finds the celebrated Bloch oscillations
with a period given by the Bloch time $T_B = 4 \pi \hbar / \lambda F_0$.
The maximum displacement of the wave packet is given by
$d = \Delta/F_0$, $\Delta$ being the width of the ground Bloch
band \cite{Hart04}. 

A directed transport of the atoms can be realized by means of a 
time-periodic driving. During one period of the driving the atoms
are first accelerated and the decelerated back to a standstill, such
that the atomic wavefunction is displaced in space but otherwise
unaffected. 
An example for directed transport in a driven optical lattice is shown
in Fig.~\ref{fig-flipped1} (b) for a 
sinusoidal driving
\be
   F(t) = F_1 \cos(2\pi t/T),
\ee
assuming a driving strength of $F_1/m = 0.42 \, {\rm m}/{\rm s}^2$
with a period of $T = 10 \, {\rm ms}$. 
The transport properties of a driven optical lattice can be calculated 
analytically within the tight-binding approximation \cite{Hart04,Klum07}.
It is found that transport is possible only for a pure ac-driving, 
while it is generally forbidden in the case of a combined dc- and ac-field 
$F(t)$ except for the case of a resonant driving $T = n T_B$.
The actual transport velocity depends on the initial state, but an
upper bound can easily be found:  
\be
  v_{\rm max} = \frac{d \Delta}{2 \hbar} 
       J_n \left( \frac{d F_1}{\hbar \omega} \right),
\ee
where $J_n$ is an ordinary Bessel function. The case of a 
pure ac-driving corresponds to $n=0$. The dispersion of a 
wave packet is generally negligible for an initially broad wave 
packet as it vanishes with $1/\sigma^4$ \cite{Klum07,Brei07,Hall10}.
This approach to controlling transport in driven optical lattices was 
experimentally demonstrated in \cite{Sias08,Zene09}.

\begin{figure}[tb]
\centering
\includegraphics[width=9cm,  angle=0]{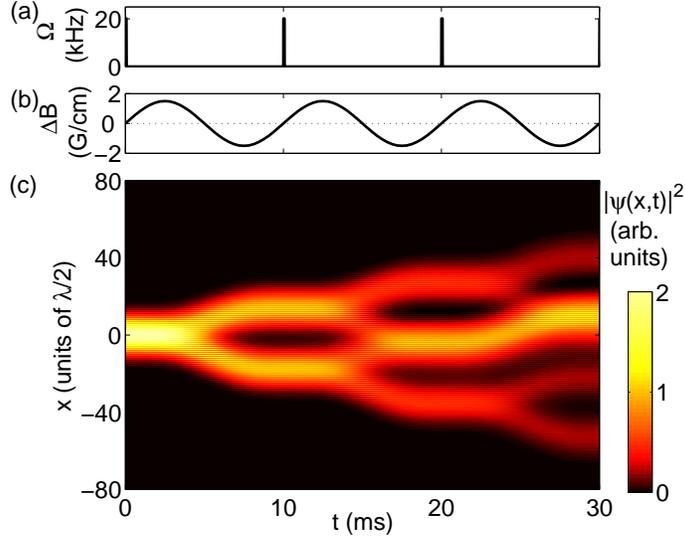}
\caption{\label{fig-qwalk21}  
(Color online)
Quantum walk of ultracold Cs atoms in a driven
magnetic gradient field.
(a) The coin operations: Rabi frequency of the microwave 
$\pi/2$-pulses inducing transitions between the two hyperfine levels.
(b) The shift operation: A driven magnetic gradient field
induces state-dependent transport.
(c) The resulting dynamics of the total atomic density
$|\psi_{\uparrow}(x,t)|^2 + |\psi_{\downarrow}(x,t)|^2$.
Parameters are given in the text.
}
\end{figure}

\section{Quantum Walks}

Now consider the dynamics of a spinor atom, where the atoms
experience a different field strength depending on their initial
state. Using Bloch oscillations and the directed transport described 
above, it is possible to entangle the position and the internal
degrees of freedom. This provides the basis for the implementation
of a quantum walk or a Dirac quantum simulator.

One possibility to induce a state dependent transport of
spinor atoms is the use of a magnetic gradient field. In particular 
we consider the dynamics of ultracold cesium atoms with the
internal states $|\uparrow \rangle =  | F=4, m_F = 4 \rangle$ 
and $|\downarrow \rangle =  | F=3, m_F = 3\rangle$
\cite{Gorl03}. The magnetic moments of these two internal states 
are opposite, such that they move into different
directions in a magnetic gradient field. The effective
potential is given by
\be
  V_{m_F}(x) =    g_F m_F \mu_B \, \Delta B \, x,
\ee
assuming a linear magnetic gradient field $B_z(x) = \Delta B \, x$.
The effective Land\'e factor of the two internal states is given
by $g_{F=4} = 1/4$ and $g_{F=3} = -1/4$ \cite{Arim77}
and $\mu_B = 9.274 \times 10^{-24} {\rm J/T}$ denotes 
the Bohr magneton. The magnetic gradient field can also
be varied in time to control the atomic motion \cite{Hall10}.
Furthermore, we assume that the internal state of the atoms
can be manipulated coherently by resonant microwave pulses
with Rabi frequency $\Omega(t)$ \cite{Fors09}.
The dynamics of the atomic state 
$(\psi_{\uparrow}(x,t),\psi_{\downarrow}(x,t))$ 
is thus given by the Hamiltonian
\bea
  \hat H(t)
     &=& \frac{-\hbar^2}{2m} \frac{\partial^2 }{\partial x^2} 
        + V_0 \cos^2(k_0 x) \nn \\
       && \quad  +  \left( \begin{array}{c c}
     V_{\uparrow}(x) & 0  \\
     0 & V_{\downarrow}(x) \end{array} \right) 
      + \frac{\hbar \Omega(t)}{2} \hat \sigma_x.
\eea

\begin{figure}[tb]
\centering
\includegraphics[width=9cm,  angle=0]{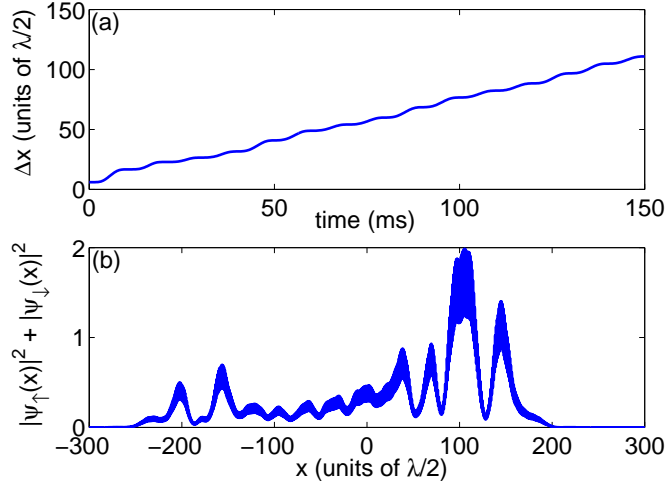}
\caption{\label{fig-qwalk22}  
(Color online)
Characterization of a Bloch quantum walk:
(a) Increase of the standard deviation 
$\Delta x = \sqrt{\langle x^2\rangle- \langle x \rangle^2}$.
(b) The total atomic total atomic population
$|\psi_{\uparrow}(x,t)|^2 + |\psi_{\downarrow}(x,t)|^2$
after 15 driving periods. 
The parameters are given in the text.
}
\end{figure}

A coherent quantum walker can be implemented by a
sinusoidal driving of the gradient field $\Delta B(t) = \Delta B_0
\times \sin(2\pi t/T)$. During one period, the 
atom are first accelerated and then decelerated  back to a
standstill as illustrated in Fig.~\ref{fig-flipped1} (b). 
However, the transport direction depends on the direction 
of the external field, which is opposite for both internal states, 
thus realizing an effective quantum walk. The coin operation is 
then realized by a $\pi/2$ pulse coupling the two hyperfine 
levels after integer multiples of the driving time $T$. 
An example of a quantum walk is shown in Fig.~\ref{fig-qwalk21} 
(c),  where we have plotted the dynamics of the complete atomic density 
$|\psi_{\uparrow}(x,t)|^2 + |\psi_{\downarrow}(x,t)|^2$. The time-dependence 
of the magnetic gradient field and the microwave pulses is illustrated in 
the upper panels of the figure. Initially, all atoms are assumed to be in 
the internal state $\ket{\downarrow} \otimes \psi_0(x)$,
where the spatial wavefunction $\psi_0(x)$ is a Bloch state with $\kappa = 0$,
weighted with a Gaussian envelope of width $\sigma = 3 \, \lambda$.
The depth of the optical lattice is given by $V_0 = E_R$.

It has to be noted that the magnitude of the force is different for the 
two spin states leading to a trivial overall displacement of the atomic 
density pattern. Furthermore this introduces a phase 
shift between the two spinor components, which has more
severe consequences.
This phase shift can be avoided by choosing internal states with
the same $m_F$, i.e $\ket{\uparrow} = \ket{4,3}$ instead of
$\ket{4,4}$, or it can be compensated actively by a static
magnetic field.  

\begin{figure}[tb]
\centering
\includegraphics[width=9cm,  angle=0]{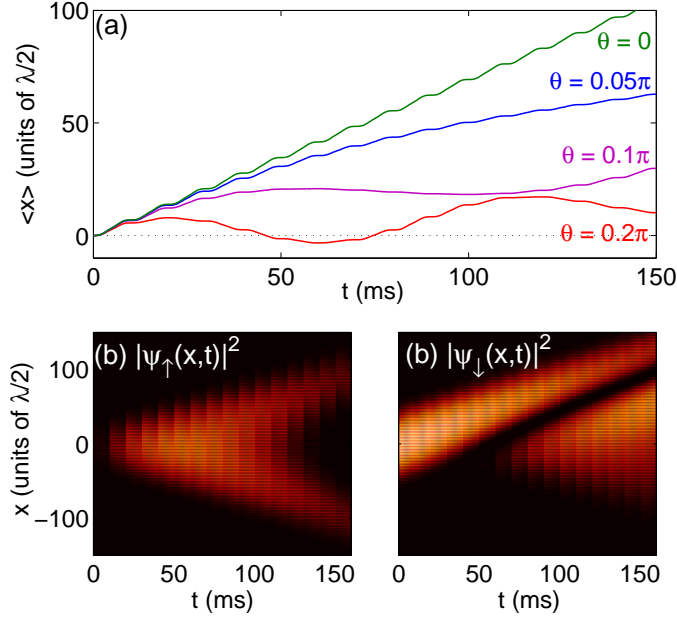}
\caption{\label{fig-zitter}  
(Color online)
Zitterbewegung of spinor atoms in a driven magnetic gradient
field. (a) Evolution of the position expectation value for different
values of the rotation angle $\theta$.
(b,c) Evolution of the probability densities $|\psi_{\uparrow}(x,t)|^2$ 
and $|\psi_{\downarrow}(x,t)|^2$ for $\theta = 0.1 \pi$.
The remaining parameters are given in the text. 
}
\end{figure}

\section{Dirac Dynamics}

For weak driving, the dynamics of a quantum walker is equivalent
to that of a Dirac particle in 1+1 dimension \cite{Stra06}, which can 
be seen as follows. During one period $T$, the atoms are first 
displaced depending on their internal state and then the internal 
state is rotated by an angle $\theta$.
Assuming that the internal rotation is fast compared to $T$, the
dynamics is given by the evolution operator 
$\hat U(nT) = (\hat U_2 \hat U_1)^n$
with  
\be
   \hat U_1 = \left( \begin{array}{c c}
       \hat D_d & 0 \\ 0 & \hat D_{-d} \end{array}  \right)
     \quad \mbox{and} \quad  
   \hat U_2 = \exp(- i \theta \hat \sigma_x/2), 
   \label{eqn-u12}
\ee
where $\hat D_d$ denotes the translation operator over a distance
$d$.
In the limit of weak driving, i.e small values of $d$ and $\theta$, 
the dynamics can be described by an effective Hamiltonian,
 $\hat U(nT) \approx \exp(-i \hat H_{\rm eff} n)$, which is given by 
\be
   \hat H_{\rm eff} = d \hat p \hat \sigma_z + \frac{\theta}{2} \hat \sigma_x.
   \label{eqn-ham-dirac}
\ee
This is just the Dirac Hamiltonian with an effective mass
$m = \theta/2d$ and the momentum operator $\hat p = -i \partial/\partial x$.

\begin{figure}[tb]
\centering
\includegraphics[width=9.5cm,  angle=0]{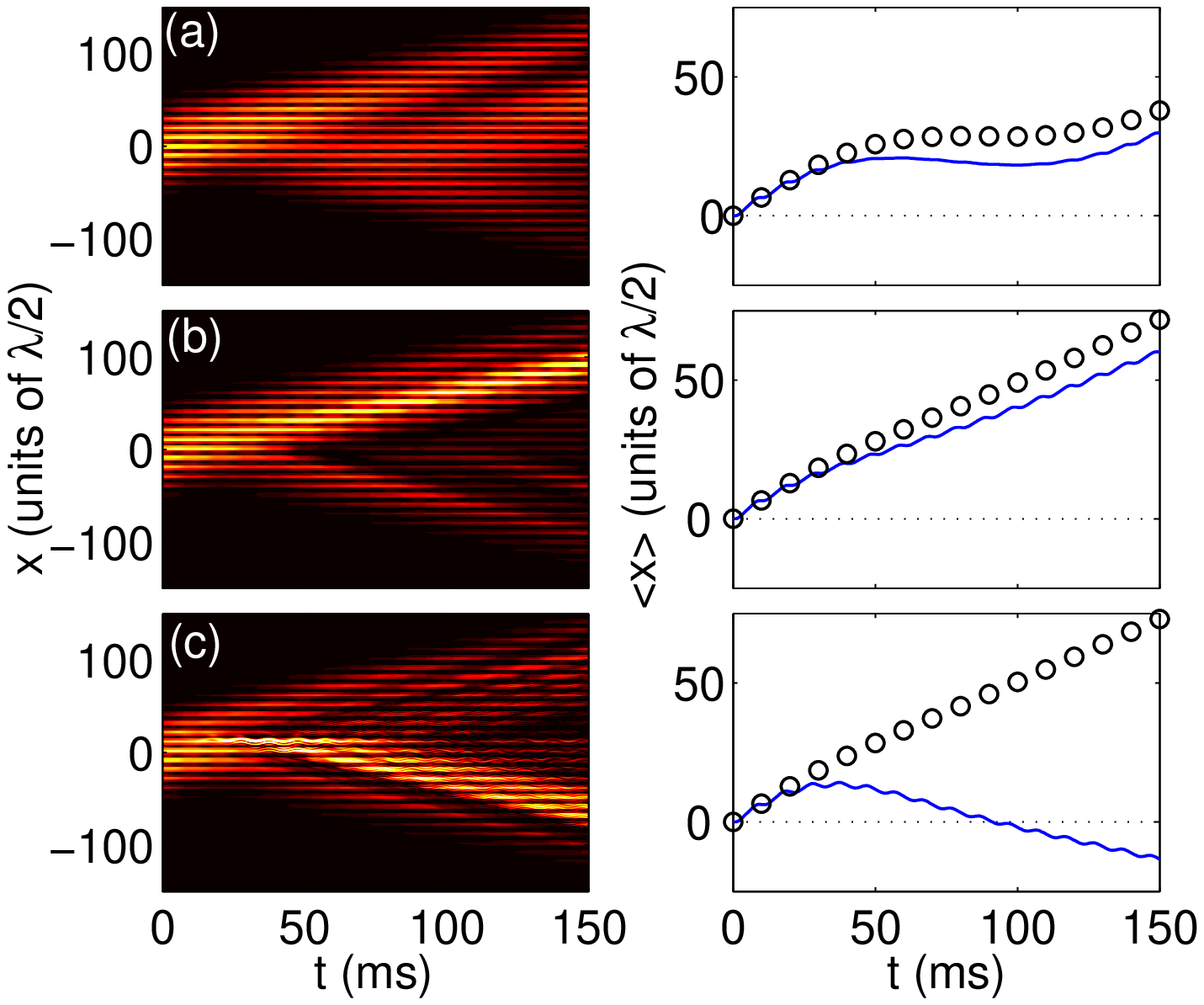}
\includegraphics[width=8.5cm,  angle=0]{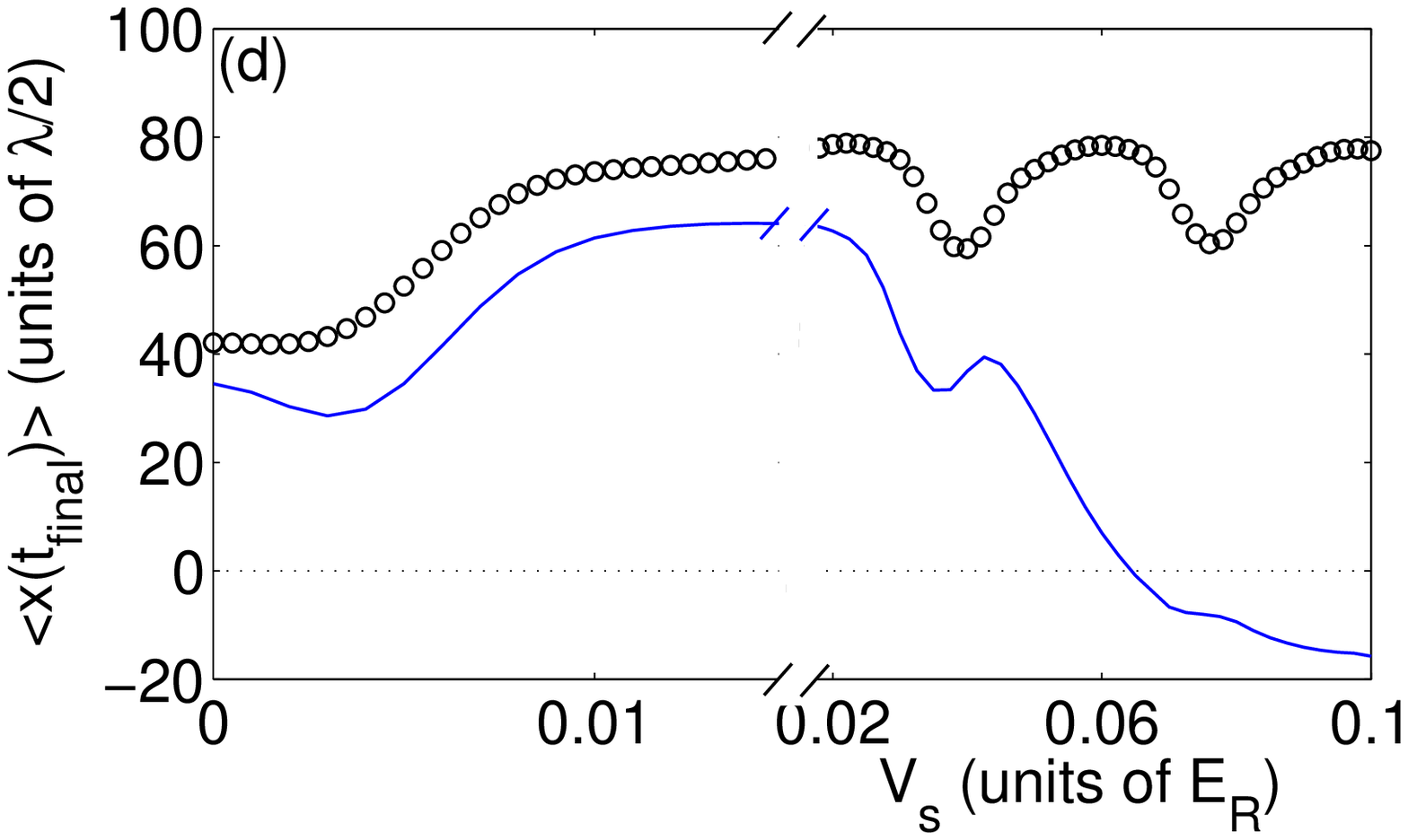}
\caption{\label{fig-klein}  
(Color online)
Dynamics of spinor atoms in the Dirac regime with an 
additional potential step. (a-c) Evolution of the atomic
density (left) and the position expectation value 
$\langle x(t) \rangle$ (right) for different heights of the
potential step: $V_{\rm s}/E_R = 0$ (a), $0.015$ (b)
and $0.1$ (c). 
(d) Position expectation value $\langle x(t) \rangle$ after a
fixed propagation time $t_{\rm final} = 150 \, {\rm ms}$ as a 
function of the potential step $V_{\rm s}$.
Results of a full quantum simulation (solid blue line) are compared 
to the effective discrete Dirac approximation (open circles).
The remaining parameters are the same as in Fig.~\ref{fig-zitter} 
with $\theta = 0.1 \, \pi$. 
}
\end{figure}

According to these results, ultracold spinor atoms show an effective
relativistic dynamics if the displacement $d$ and the rotation angle
$\theta$ are reduced. This is simply realized by increasing the lattice 
depth $V_0$ and decreasing the strength of the microwave driving 
field. An example is shown in Fig.~\ref{fig-zitter} for atoms initially
prepared in the internal state $\ket{\downarrow} = \ket{3,3}$ with
a width $\sigma = 10 \, \lambda$. The 
optical lattice is now deeper as in the previous examples, $V_0 = 5 E_R$.
Transitions to the other spin state $\ket{\uparrow} = \ket{4,3}$ are
induced at integer multiples of the period $T_{\rm drive}$ with 
a variable rotation angle $\theta = \Omega t_{\rm pulse}$. 
For $\theta = 0$, the internal state does not change such that the 
atoms are steadily transported by the time-periodic external force.
Increasing $\theta$ and thus the effective mass of the Dirac particle
leads to a Zitterbewegung of the atoms, which is illustrated in 
Fig.~\ref{fig-zitter} (a). As expected, the period of the oscillations 
decreases with the effective mass $m$. Microscopically, the
oscillations of the position expectation value $\langle x(t) \rangle$
result from the interference of the two components of the atomic 
state, which are shown in Fig.~\ref{fig-zitter} (b,c)

One of the most surprising predictions of Dirac theory is
the Klein paradoxon: Relativistic particles are not repelled
by a strong repulsive potential, but perfectly transmitted if the 
height of the potential step exceeds $2mc^2$ \cite{Thal92}.
Signatures of this effect can be readily explored with ultracold
atoms in driven optical lattices. We consider a situation where
an additional blue detuned tophat laser beam is focused aside
of the atom cloud. We model the induced optical dipole potential 
by a tanh profile such that the total optical potential is given by
\bea
  V(x) &=& V_{\rm lattice}(x) + V_{\rm step}(x)  \\
   &=& V_0 \cos^2(k_0 x) + \frac{V_{\rm s}}{2}  (\tanh((x-x_s)/w) + 1). \nn 
\eea
If the additional potential is weak enough, the directed transport 
mechanism remains mostly unaffected. However, the atomic wave 
packets accumulate a dynamical phase depending on their position 
which is equivalent to a potential in the effective Dirac hamiltonian 
(\ref{eqn-ham-dirac}). In this regime, the dynamics is described 
by the discrete evolution operator 
\bea
    \hat U(nT) &\approx& (\hat U_3 \hat U_2 \hat U_1)^n \quad \mbox{with} \nn \\
    \hat U_3(x) &=& \exp(-i V_{\rm step}(x) \,  T/\hbar)
    \label{eqn-dirac-pot}
\eea
and $\hat U_{1,2}$ given in Eqn.~(\ref{eqn-u12}). The dynamics
in the presence of an additional potential is shown in 
Fig.~\ref{fig-klein}, comparing a full simulation to the Dirac 
approximation (\ref{eqn-dirac-pot}). A good agreement is observed 
for weak  potential steps, $V_{\rm s} \apprle 0.02 \, E_R$. 

\begin{figure}[tb]
\centering
\includegraphics[width=9cm,  angle=0]{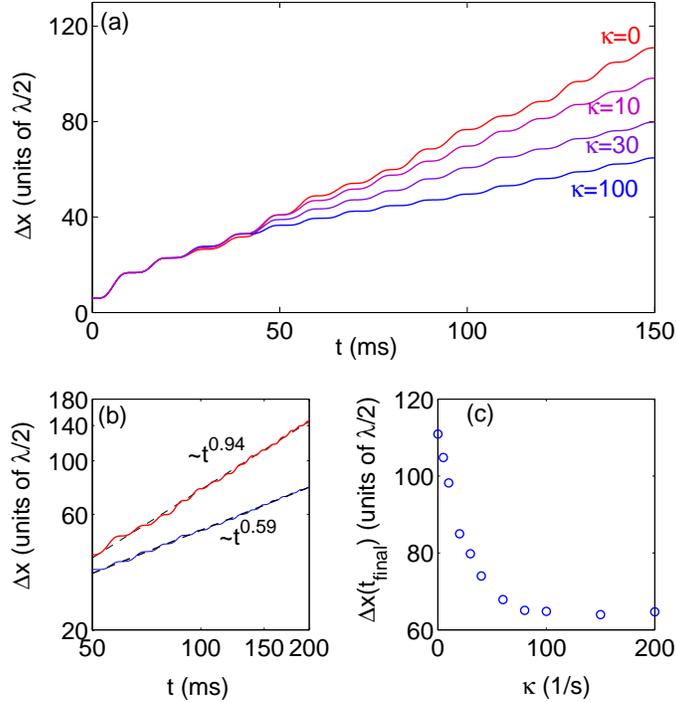}
\caption{\label{fig-qwalk-dec}  
(Color online)
Quantum walk in the presence of decoherence.
(a) Increase of the standard deviation $\Delta x(t)$
for different values of the dephasing rate $\kappa$.
(b) The standard deviation in a loglog plot showing a different 
scaling $\Delta x(t) \sim t^{\alpha}$ in the quantum regime
($\kappa = 0$) and in the classical regime
($\kappa = 100 \, {\rm s}^{-1}$). The dashed black line is
a linear fit with the result $\alpha = 0.94$
and $\alpha = 0.59$, respectively.
(c) The standard deviation $\Delta x(t_{\rm final})$ after
a fixed time $t_{\rm final} = 150 \, {\rm ms}$
as a function of $\kappa$.
}
\end{figure}

A pronounced signature of Klein tunneling observed in 
Fig.~\ref{fig-klein} is that the repulsive potential step 
{\em enhances} the transmission of the atomic matter 
wave. Without the potential, the atoms exibit a Zitterbewegung 
as discussed above. If a weak additional potential step is 
included, e.g. $V_{\rm s} = 0.015 \, E_R$, the oscillating 
motion is gone and the atoms move steadily above the barrier.
An even stronger potential, however, suppresses the directed
transport underlying the Dirac quantum simulator. The effective
description by the Dirac hamiltonian (\ref{eqn-ham-dirac}) is no 
longer appropriate and the atoms are reflected just like ordinary
Schr\"odinger particles. In this regime the Dirac approximation
clearly deviates from the exact simulation results.

\section{Decoherence}

Real experiments with spinor atoms are limited by a loss of coherence
due a coupling to the environment. The schemes discussed above make
use on the opposite magnetic moment of the two hyperfine states 
$\ket{\uparrow}$ and $\ket{\downarrow}$ to transport the atoms in 
different directions. However, this makes the atomic state vulnerable 
to dephasing caused by magnetic field fluctuations \cite{Kars09},
while spatial coherence can be conserved over seconds \cite{Gust08}.
To explore the effects of decoherence, we simulate the dynamics 
in the presence of pure dephasing described by the Master equation 
\be
\frac{\partial \hat \rho}{\partial t} = -i [\hat H,\hat \rho] 
    - \frac{\kappa}{2} \left( \hat \sigma_z^2 \hat \rho
        + \hat \rho \hat \sigma_z^2 
         - 2 \hat \sigma_z \hat \rho \hat \sigma_z \right).
\ee
For the numerical calculations we use the quantum jump method 
\cite{Dali92}, averaging over 100 trajectories.

\begin{figure}[tb]
\centering
\includegraphics[width=9cm,  angle=0]{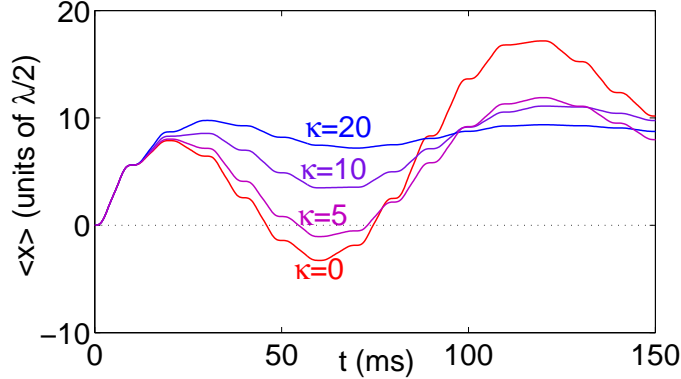}
\caption{\label{fig-zitter-dec}  
(Color online)
Atomic {\it Zitterbewegung} in the presence of decoherence.
Shown is the position expectation value $\langle x(t) \rangle$
for different values of the dephasing rate $\kappa$. Parameters
are the same as in Fig.~\ref{fig-zitter} with $\theta = 0.2 \, \pi$.  
}
\end{figure}

Decoherence of the atomic states essentially turns a quantum
random walk into a classical one, which is confirmed by the 
simulations shown in Fig.~\ref{fig-qwalk-dec}. After a short
initial period the standard deviation $\Delta x$ grows much
slower in the the presence of dephasing. The linear increase
$\Delta x(t) \sim t$ gradually changes into the classical 
diffusion law $\Delta x(t) \sim t^{1/2}$ if $\kappa$ is increased.
The different scaling is analyzed quantitatively in part (b),
where $\Delta x(t)$ is plotted on a loglog scale together with
a linear fit, omitting the initial expansion of the atomic matter 
wave. The different slope of the two curves directly reveals 
the different diffusion exponents. The fit yields an exponent
of $\alpha = 0.94$ and $\alpha = 0.59$ for $\kappa = 0$ 
and $\kappa = 100 \, {\rm s}^{-1}$, respectively, which is in 
good agreement with the theoretical expectation $\alpha = 1$
for a quantum and $\alpha = 1/2$ for a classical random walk.
A quantitative analysis of the decoherence process is
provided in Fig.~\ref{fig-qwalk-dec} (c), where the 
standard deviation $\Delta x$ after a fixed 
propagation time $t_{\rm final} = 150 \, {\rm ms}$ is plotted
as function of $\kappa$. One observes that $\Delta x$
decreases with $\kappa$ until the random walk becomes
completely classical at  $\kappa \approx 80 \, {\rm s}^{-1}$.

Also the Dirac quantum simulator discussed in the previous 
section is vulnerable to decoherence of the atomic hyperfine 
states. Figure \ref{fig-zitter-dec} shows an example of the atomic 
{\it Zitterbewegung} for the same parameters as in 
Fig.~\ref{fig-zitter} for different values of the dephasing rate 
$\kappa$. The oscillations of the mean position of the atoms 
are much less pronounced for $\kappa > 0$. We find that the 
dephasing rate must not exceed $20 \, {\rm s}^{-1}$ to observe 
a clear signature of a {\it Zitterbewegung}.
For stronger dephasing no interference effects and thus
no oscillations are visible.
  
\section{Conclusion}

In the present paper we have explored the rich dynamics of
spinor atoms in tilted and driven optical lattices. These systems
are nowadays routinely realized experimentally and can be controlled 
with a high accuracy. We have discussed how a state-dependent transport
can be realized by a periodic driving and how this can be used
to implement a quantum simulator. A quantum random walk is
realized when transitions between the internal states are driven
by microwave or Raman pulses. 
In the continuum limit of weak driving, the dynamics is given by
the Dirac equation in 1+1 dimension. Thus it is possible to 
investigate relativistic effects such as Zitterbewegung or Klein
tunneling in a well controllable laboratory experiment.

All elements needed for the implementation of the proposed
quantum simulator have been demonstrated quite recently 
such that an experimental realization is possible with current
technology. A major obstacle can be decoherence of the 
internal atomic state.
On the other hand, decoherence processes offer the possibility 
to experimentally study the quantum classical transition 
for a Dirac particle. 

\begin{acknowledgments}

This work has been supported by the German Research Foundation (DFG) 
through the research fellowship program (grant no WI 3415/1).
I thank F.~Trimborn and H.~J. Korsch for valuable comments.

\end{acknowledgments}

% --- Literatur -------------------------------------------------------------------

\end{document}